\documentclass[aps,pra,twocolumn,superscriptaddress,showpacs,showkeys,amsmath,amssymb]{revtex4-1}

\usepackage{amssymb}
\usepackage{mathrsfs}
\usepackage{latexsym}
\usepackage{dcolumn}
\usepackage{color}
\usepackage{amsmath}
\usepackage{accents}
\usepackage{bm}
\usepackage{amsfonts}
\usepackage[english]{babel}
\usepackage{graphicx}
\usepackage[font=footnotesize,figurewithin=none]{caption}
\usepackage{subcaption}
\usepackage{floatrow}
\usepackage[toc,page,title]{appendix}

\begin{document}

\title{Implementation of a two-qubit state by an auxiliary qubit\\ on the three-spin system}

\author{A. R. Kuzmak}
\email{andrijkuzmak@gmail.com}
\affiliation{Department for Theoretical Physics, Ivan Franko National University of Lviv,\\ 12 Drahomanov St., Lviv, UA-79005, Ukraine}

\date{\today}
\begin{abstract}
The method for preparation of a two-qubit state on two spins-$1/2$ that mutually interact through an auxiliary spin is proposed.
The essence of the method is that, initially, the three spins evolve under the action of an external magnetic field during
a predefined period of time. Then, the auxiliary spin is measured by a monochromatic electromagnetic radiation that allows
obtaining a certain state of the remaining spins. We study the entanglement of this state and obtain
the condition for achieving the maximally entangled state. The implementation of the method on the physical system of nuclear spins
of xenon difluoride is described. As a results, the conditions which allow preparing the maximally entangled state on this system are obtained.
\end{abstract}

\maketitle

\section{Introduction \label{sec1}}

An important problem of quantum information is finding the methods for preparation of predefined quantum states that in turn allow to execute
certain quantum algorithms. Quantum algorithms consist of a circuit of unitary operators which provide the transformation
of a quantum register composed of quantum bits (qubits) \cite{Kitaev2002,Krokhmalskii2004,Barenco1995}. These unitary operators are also called
quantum gates. The qubits should be prepared on two-level quantum systems which are well isolated from their environment,
for providing a high degree of quantum coherence. The evolution of such systems is controlled by devices, which provide the unitary transformation.
The states prepared on such systems should be measured with high fidelity. The systems that satisfy these conditions were suggested in many papers:
spins of electrons and nuclei of atoms
\cite{qdots1,qdots2,qdots3,phosphorus3,scc,semcond1,semcond2,semcond3,semcond4,semcond5,semcond6,phosphorus1,phosphorus2,phosphorus4,phosphorus6,phosphorus7,phosphorus8,HinSi,BiinSi,LNT,Christle2017,Nichol2017,Harvey-Collard2018,Nagy2019},
superconducting circuits
\cite{supcond1,supcond2,supcond3,supcond4}, trapped ions \cite{SchrodCat1,EQSSTI,ITQLLWR,SchrodCat2,ETDIITIQSHI,QSDEGHTI,QSWTI} and ultracold atoms
\cite{opticallattice1,opticallattice18,opticallattice5,opticallattice19,opticallattice6,opticallattice7,opticallattice9}, etc. Also the implementation
of qubits on different physical systems is considered in reviews \cite{Ladd2010,Buluta2011}.

Depending on the type of a physical system there are different approaches that allow to control their evolution. The evolution of the electron and nucleus
spins of atoms is provided by the interaction between them, and the interaction with external magnetic field and electromagnetic pulses.
The technique which allows to drive and measure the state of such systems is called the spin resonance technique \cite{srtech,Vandersypen2004}.
Using this technique the authors of paper \cite{Mehring2003} reported the preparation of a two-qubit entangled state between the electron and
nuclear spins in a molecular single crystal. Entanglement was measured by using a special entanglement detector sequence based on a unitary
back transformation including phase rotation. In paper \cite{Kuzmak2014}, a two-step method for the preparation of an arbitrary quantum state
on the two-spin system with isotropic Heisenberg interaction was proposed. This method is based on the interaction between the spins and
the application of an individual magnetic field to them. However, the experimental technics do not allow to control individually each spin
with help of the magnetic field. Therefore, a simplified version of this method was considered on the physical system of an atom
having with a nuclear spin $1/2$ and one valence electron. Using this method, the conditions for implementation of different quantum gates on
the physical system of ultracold atoms in optical lattice was obtained in paper \cite{Kuzmak2018}. Experimental realization of
a long-distance entangled state between spins in antiferromagnetic quantum spin chains was described in paper
\cite{Sahling2015}. The authors experimentally showed that unpaired separated spins entangled through a collection of spin singlets
made up of antiferromagnetic spin-$1/2$ chains. The implementation of quantum states on various physical systems was widely studied both
theoretically and experimentally \cite{Nichol2017,Harvey-Collard2018,Nagy2019,twosqg1,twosqg2,twosqg3,twosqg4,twosqg5,twosqg6,twosqg7,Zu2014,Ashhab2012}.

We propose the method for preparation of a two-qubit state on basic two spins which mutually interact through the auxiliary spin (Sec. \ref{sec2}).
At the first step, under the action of the external magnetic field we consider the evolution of three spins having started from the factorized state.
During the evolution, the system achieves a time-dependent state which at a predefined moment of time should be measured.
We propose to measure the state of an auxiliary spin by the monochromatic electromagnetic radiation. This allows
to obtain the predefined state of the basic spins with certain probability (Sec. \ref{sec3}). Unlike the methods mentioned in previous
papers \cite{Mehring2003,Kuzmak2014}, where the whole system should be measured, we suggest to measure the auxiliary spin that allows to
prepare the required state of the system. This fact allows to simplify the measurement process. Unfortunately, the problem of
controlling each spin individually persists. Also we study the entanglement of the state prepared by our method
and obtain the conditions for achieving the maximally entangled state (Sec. \ref{sec4}). Finally,
we describe the implementation of the method on the physical system of nuclear spins of xenon difluoride (Sec. \ref{sec5}).
As a result, we calculate the conditions which allow to prepare the maximally entangled state on this system. Conclusions are presented in Sec. \ref{sec6}.

\section{Method \label{sec2}}

\begin{figure}
\includegraphics[width=0.5\textwidth]{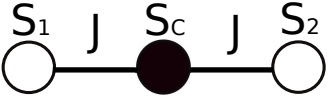}
\caption{Model of the system: spins $S_1$ and $S_2$ mutually interact through the spin $S_c$ with the coupling constant $J$.}
\label{model}
\end{figure}

We propose the method for preparation of a two-qubit quantum state on two spins using an auxiliary spin. The system, which we consider
for the implementation of this method, consists of three spins-$1/2$: the two spins $S_1$ and $S_2$ as the basic qubits which mutually interact through
the central spin $S_c$ as the auxiliary qubit (Fig. \ref{model}). The interaction between spins is described by the isotropic Heisenberg Hamiltonian
\begin{eqnarray}
H_{s}=J{\bf S}_c({\bf S}_1+{\bf S}_2),
\label{sysham}
\end{eqnarray}
where $J$ is the coupling constant and ${\bf S}_i$ is the operator which defines the $i$th spin.

The method consists of three steps.
At the first step, we prepare the initial state of the system. Then, the interactions between spins and external magnetic field
provide the evolution of the system. Also, the presence of a magnetic field is required to provide the measure of the auxiliary spin.
Because it splits the degenerate energy levels of system that allows to measure them.
Finally, at the given moment of time we make a measurement of the central spin that allows
to interrupt the interaction between basic spins and to maintain their quantum state. Before considering this method in detail, let us
rewrite Hamiltonian (\ref{sysham}) with magnetic field in a more convenient form that allows us to simplify the future calculations.

Hamiltonian (\ref{sysham}) with magnetic field can be represented as follows
\begin{eqnarray}
&&H=\frac{J}{2}{\bf S}^2+B\hbar\gamma S^z-\frac{J}{2}\left({\bf S}_{12}^2+{\bf S}_c^2\right)\nonumber\\
&&+B\hbar\left(\gamma_c-\gamma\right)S_c^z,
\label{sysham2}
\end{eqnarray}
where ${\bf S}={\bf S}_c+{\bf S}_1+{\bf S}_2$ is the operator of total spin, $S^z=S_c^z+S_1^z+S_2^z$ is the $z$ component of this operator,
${\bf S}_{12}={\bf S}_1+{\bf S}_2$ is the operator of sum of the $S_1$ and $S_2$ spins, $B$ is the value of the magnetic field, $\gamma_c$ and $\gamma$
are the gyromagnetic ratios of the auxiliary and basic spins, respectively. The eigenvalues and eigenstates of Hamiltonian (\ref{sysham2})
are presented in Appendix \ref{appa}. It is worth noting that the interections between spins are isotropic which in turn means
that the Hamiltonian of the system does not depend on the orientation of an external magnetic field.

So, let us consider the evolution of the spin system described by Hamiltonian (\ref{sysham2}) having started from the initial state
$\vert\psi_I\rangle=\vert\uparrow\uparrow\downarrow\rangle$. Here, the first vector defines the state of the auxiliary spin and the remaining
two vectors describe the states of basic spins. This state can be prepared on experiment if the spins
are placed in a strong magnetic field, such that the interaction between the field and spins is much higher than the interaction between spins.
Then the initial state $\vert\uparrow\uparrow\downarrow\rangle$ is the eigenstate of such a system. To simplify the calculations
we decompose the initial state in the basis of the eigenstates of Hamiltonian (\ref{sysham2}). Using results from Appendix \ref{appa}, we obtain
the following expression
\begin{eqnarray}
&&\vert\psi_I\rangle=\frac{1}{\sqrt{2}}\left[\vert\psi^{(2)}\rangle-\sqrt{\frac{a_-}{a_--a_+}}\vert\psi^{(3)}\rangle \right.\nonumber\\
&&\left.+\sqrt{\frac{a_+}{a_+-a_-}}\vert\psi^{(4)}\rangle\right],
\label{instate}
\end{eqnarray}
where $a_{\pm}$ are dimensionless values which are determined by expression (\ref{dimlessval}).
The evolution of the system is described by the Schr\"odinger equation and can be expressed as the unitary transformation
\begin{eqnarray}
\vert\psi(t)\rangle=e^{-iHt/\hbar}\vert\psi_I\rangle.
\label{untransf}
\end{eqnarray}
The evolution of the three spins defined by Hamiltonian (\ref{sysham}) having started from state (\ref{instate}) takes the form
\begin{eqnarray}
&&\vert\psi(t)\rangle=\frac{1}{\sqrt{2}}\left[e^{-iE^{(2)}t/\hbar}\vert\psi^{(2)}\rangle\right.\nonumber\\
&&-\sqrt{\frac{a_-}{a_--a_+}}e^{-iE^{(3)}t/\hbar}\vert\psi^{(3)}\rangle\nonumber\\
&&\left.+\sqrt{\frac{a_+}{a_+-a_-}}e^{-iE^{(4)}t/\hbar}\vert\psi^{(4)}\rangle\right],
\label{evolution}
\end{eqnarray}
where $E^{(i)}$ are the eigenvalues of the system which correspond to eigenstates $\vert\psi^{(i)}\rangle$ (see (\ref{hamevandes}) in Appendix \ref{appa}).
Using the explicit form of these eigenvalues and making some simplifications we obtain that the time-dependent state of the system is defined by the
following expression
\begin{eqnarray}
&&\vert\psi(t)\rangle=\frac{e^{-iB\gamma t/2}}{2}\nonumber\\
&&\times\left[\left(\vert z\vert e^{i(\phi+Jt/(4\hbar))}+e^{-iB(\gamma_c-\gamma)t/2}\right)\vert\uparrow\uparrow\downarrow\rangle\right.\nonumber\\
&&\left.+\left(\vert z\vert e^{i(\phi+Jt/(4\hbar))}-e^{-iB(\gamma_c-\gamma)t/2}\right)\vert\uparrow\downarrow\uparrow\rangle\right]\nonumber\\
&&+\frac{2ie^{i\left(Jt/(4\hbar)-B\gamma t/2\right)}}{a_+-a_-}\sin\left(\frac{Jt}{4\hbar}(a_+-a_-)\right)\vert\downarrow\uparrow\uparrow\rangle,
\label{timedepstate}
\end{eqnarray}
where we introduce the notation of the following complex number
\begin{eqnarray}
z=\cos\left(\frac{Jt}{4\hbar}(a_+-a_-)\right)+i\frac{a_++a_-}{a_+-a_-}\sin\left(\frac{Jt}{4\hbar}(a_+-a_-)\right)\nonumber
\end{eqnarray}
with the modulus and argument
\begin{eqnarray}
&&\vert z\vert=\sqrt{1-\frac{8\sin^2\left(\frac{Jt}{4\hbar}(a_+-a_-)\right)}{(a_+-a_-)^2}},\nonumber\\
&&\phi=\arctan\left(\frac{a_++a_-}{a_+-a_-}\tan\left(\frac{Jt}{4\hbar}(a_+-a_-)\right)\right),\nonumber
\end{eqnarray}
respectively.

Finally, at the moment of time $t_f$ we measure the auxiliary spin that allows to obtain a certain state of the remaining spins.
From the analysis of state (\ref{timedepstate}) it follows that if the auxiliary spin takes the state $\vert\downarrow\rangle$ then the remaining spins
are defined by state $\vert\uparrow\uparrow\rangle$. This state is a factorized state and is of no interest because it can be prepared in
the same way as the initial state. Otherwise, if the state of the auxiliary spin takes the form $\vert\uparrow\rangle$ then the state of
the basic spins is defined by the following expression
\begin{eqnarray}
&&\vert\psi_f\rangle=\frac{1}{\sqrt{2(1+\vert z_f\vert^2)}}\nonumber\\
&&\times\left[\left(\vert z_f\vert e^{i(\phi_f+Jt_f/(4\hbar))}+e^{-iB(\gamma_c-\gamma)t_f/2}\right)\vert\uparrow\downarrow\rangle\right.\nonumber\\
&&\left.+\left(\vert z_f\vert e^{i(\phi_f+Jt_f/(4\hbar))}-e^{-iB(\gamma_c-\gamma)t_f/2}\right)\vert\downarrow\uparrow\rangle\right],
\label{timedepstate2}
\end{eqnarray}
where $\vert z_f\vert$ and $\phi_f$ correspond to the moment of time $t_f$. It is worth noting that the external magnetic field $B$ and effective
magnetic field $J/2$, which appears due to interaction of basic spins with the measured auxiliary spin, do not influence
state (\ref{timedepstate2}). Also, it is important to note that the local manipulations with each spins separately allow achieving an arbitrary
state of two qubits \cite{Kuzmak2014}. Let us describe the measurement process in detail.

\section{Measurement of the auxiliary spin \label{sec3}}

In this section, we describe the measurement process of the central spin and, as a result, present the physical quantities which should be
measured experimentally to obtain a certain state of system. Measurement of the state of spin system is provided by the interaction of it with
electromagnetic field and magnetic field \cite{srtech,Vandersypen2004}. For this purpose, we consider the interaction of spin system defined by Hamiltonian
(\ref{sysham2}) with monochromatic electromagnetic radiation with frequency $\omega$, wave vector directed along
the $x$ axis ${\bf k}=(k,0,0)$ and polarization vector located in the plane perpendicular to the direction of propagation of the wave
${\bf e}=(0,e_y,e_z)$. We assume that this interaction occurs after the evolution of the system at the moment of time $t_f$.
The spin system interacts only with the magnetic component of the field. Therefore, the Hamiltonian of this interaction has the form
\begin{eqnarray}
&&H_{int}=B_{int}\hbar\left[\left(\gamma_cS_c^z+\gamma(S_1^z+S_2^z)\right)e_y\right.\nonumber\\
&&\left.-\left(\gamma_cS_c^y+\gamma(S_1^y+S_2^y)\right)e_z\right]i\left(e^{ikx}b-e^{-ikx}b^+\right),
\label{haminteraction}
\end{eqnarray}
where $B_{int}=k\sqrt{2\pi c^2\hbar/(\omega V)}$ is proportional to the value of the magnetic component of electromagnetic field, $b^+$ and $b$ are the
creation and annihilation operators of the photon with frequency $\omega$ and wave vector $k$. Here $c$ is the speed of light in the vacuum,
$V$ is the volume where the electromagnetic field is located. We assume that the spin system is placed in a homogeneous field.
We consider the interaction of spin system (\ref{sysham2}) with electromagnetic field in the first order of the perturbation theory, which
describes the absorption and emission of the photons by system. It is easy to see that to describe these processes for the auxiliary spin it is enough to
calculate the corrections to the eigenstates $\vert\psi^{(2)}\rangle$ and $\vert\psi^{(5)}\rangle$ (\ref{hamevandes}).
The difference between the energy levels of these states is determined by the difference between
the energy levels of the auxiliary spin. So, projecting the state of system on these perturbed states allows to find the probabilities
of obtaining a certain state of the auxiliary spin. It is important to note that the process of measuring of the central spin is due to
an external magnetic field which removes the degeneration of the $E^{(2)}$ and $E^{(5)}$ energy levels.

As it is mentioned in the previous section, we are interested in the case of the $\vert\uparrow\rangle$ state of the auxiliary spin.
Therefore, we calculate the probabilities which correspond to the absorption and emission of the photons by this spin in the $\vert\uparrow\rangle$ state.
These probabilities allow us to obtain the intensities that correspond to these events.
So, probabilities of the emission and absorption of photon by the auxiliary spin are calculated in Appendix \ref{appb} and they have the form
\begin{eqnarray}
&&W^+=\frac{\pi c^2}{\omega V}\frac{\sin^2(\omega_{52}\tau/2)}{\omega_{52}^2}\hbar\gamma_c^2\vert e_z\vert^2k^2(N+1),\nonumber\\
&&W^-=\frac{\pi c^2}{\omega V}\frac{\sin^2(\omega_{25}\tau/2)}{\omega_{25}^2}\hbar\gamma_c^2\vert e_z\vert^2k^2N,
\label{probab}
\end{eqnarray}
respectively. Here $\tau$ is the duration of electromagnetic irradiation, $N$ is the number of photons in the environment, and
\begin{eqnarray}
&&\omega_{52}=\frac{E^{(5)}-E^{(2)}}{\hbar}+\omega,\nonumber\\
&&\omega_{25}=\frac{E^{(2)}-E^{(5)}}{\hbar}-\omega
\label{differense}
\end{eqnarray}
are the differences between the corresponding energy levels of the system in the cases of the emission and absorption of photon by the auxiliary spin, respectively.

Now we calculate the integral intensity which is defined by the difference of the emitted and absorbed intensities. This intensity has the form
\begin{eqnarray}
&&I=\frac{V\hbar}{(2\pi)^3c^3}\int_0^{\infty}d\omega\, \omega^3 \frac{W^+-W^-}{\tau}\nonumber\\
&&=\frac{\hbar^2\vert e_z\vert^2}{16\pi c^3}B^4\gamma_c^6.
\label{intensity}
\end{eqnarray}
Here we use the fact that under the integral the following replacement can be made
\begin{eqnarray}
\lim_{\tau\to\infty}\frac{\sin^2(\omega_{ij}\tau/2)}{\tau\left(\omega_{ij}/2\right)^2}=\pi\delta\left(\omega_{ij}/2\right),\nonumber
\label{deltafunc}
\end{eqnarray}
where $\delta\left(\omega_{ij}/2\right)$ is the Dirac delta function, and $k=\omega/c$. It is easy to see from expressions (\ref{probab}) and (\ref{intensity}) that
the auxiliary spin is obtained with maximum probability in the state $\vert\uparrow\rangle$ when the frequency of electromagnetic field
is equal to $\omega=B\gamma_c$ and the polarization vector is such that $\vert e_z\vert=1$.

\section{Preparation of the entangled states \label{sec4}}

Entanglement plays a crucial role in processes related to the quantum information. For instance, the quantum teleportation of a qubit state requires
preparation of a two-qubit entangled state as a quantum channel \cite{bennet1993,zeilinger1997}.
Also, the efficiency of quantum algorithms depends on the value of entanglement which the system of qubits can take during the evolution
\cite{Kitaev2002,Krokhmalskii2004,Barenco1995,giovannetti20031,giovannetti20032,zander2007,borras20081,borras20082,zhao2009}.
Therefore it is important to find the conditions for preparation of the entangled states on the system of basic spins.

\begin{figure}
\includegraphics[width=1.00\textwidth]{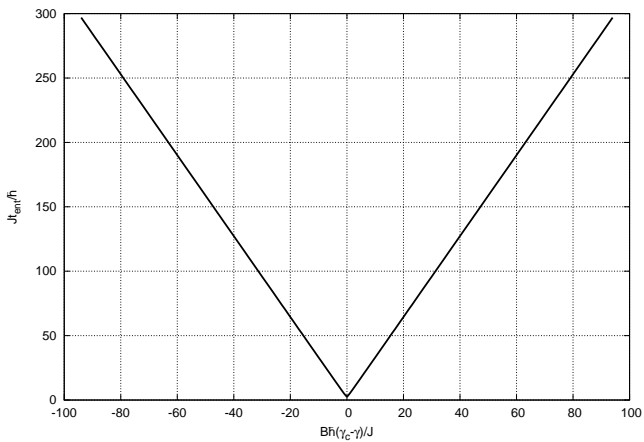}
\caption{Dependence of the moment of time at which the system achieves the maximal entangled state on the value of magnetic field.
The results are presented in the case of high magnetic field (\ref{cocncurrencecond2}).}
\label{timeevol}
\end{figure}

Using the squared concurrence as an entanglement measure, we study the entanglement of state (\ref{timedepstate2}).
For a pure state of a bipartite two-level system it is defined by expression \cite{Woo, ent1,ent2}
\begin{eqnarray}
C(\vert \psi \rangle)= 2\vert ad-bc\vert,
\label{cocncurrence}
\end{eqnarray}
where $a$, $b$, $c$ and $d$ are defined by expression
\begin{eqnarray}
\vert \psi \rangle = a\vert\uparrow\uparrow\rangle+b\vert\uparrow\downarrow\rangle+c\vert\downarrow\uparrow\rangle+d\vert\downarrow\downarrow\rangle.
\label{form12a}
\end{eqnarray}
So, for state (\ref{timedepstate2}) the concurrence takes the form
\begin{eqnarray}
&&C(\vert \psi_{f} \rangle)= \frac{1}{1+\vert z_{f}\vert^2}\left[1+\vert z_{f}\vert^4\right.\nonumber\\
&&\left.-2\vert z_{f}\vert^2\cos 2\left(\phi_{f}+\frac{Jt_{f}}{4\hbar}+\frac{B(\gamma_c-\gamma)t_f}{2}\right)\right]^{1/2}.
\label{cocncurrencef}
\end{eqnarray}
This expression takes the maximum value ($C(\vert \psi_{ent} \rangle)=1$) for condition
\begin{eqnarray}
\phi_{ent}+\frac{Jt_{ent}}{4\hbar}+\frac{B(\gamma_c-\gamma)t_{ent}}{2}=\frac{\pi}{2}+\pi n,
\label{cocncurrencecond}
\end{eqnarray}
where $\phi_{ent}$, $t_{ent}$ corresponds to the maximally entangled state $\vert\psi_{ent}\rangle$ and $n\in\mathbb{Z}$. So, for certain physical system defined by specific values of $J$, $\gamma_c$ and $\gamma$ we obtain the
relation between the moment of time $t_{ent}$ at which the system achieves the maximal entangled state and the value of magnetic field $B$.
It is easy to show that for a high magnetic field $\vert z\vert\approx 1$ and the following ratio $(a_++a_-)/(a_+-a_-)$ takes the value $\mp 1$, where upper and lower signs
correspond to a positive and negative values of $B\hbar(\gamma_c-\gamma)/J$, respectively. Then the dependence (\ref{cocncurrencecond})
can be represented as follows
\begin{eqnarray}
&&\frac{Jt_{ent}}{\hbar}\nonumber\\
&&=\frac{\pi+2\pi n}{\mp\sqrt{\frac{9}{4}+\frac{B\hbar(\gamma_c-\gamma)}{J}\left(\frac{B\hbar(\gamma_c-\gamma)}{J}+1\right)}+\frac{B\hbar(\gamma_c-\gamma)}{J}+\frac{1}{2}}.\nonumber\\
\label{cocncurrencecond2}
\end{eqnarray}
In Fig. \ref{timeevol}, we express this dependence. As we can see, the minimal value of time is achieved for the absence
of a field and it increases linearly with increasing the value of magnetic field. It is important to note that the value of time for the zero magnetic
field can be calculated from equation (\ref{cocncurrencecond}) if $B=0$. This time is determined by the expression $Jt_{ent}/\hbar\simeq 2.39$.
The egenvalues of Hamiltonian (\ref{sysham2}) become degenerate as follows $E^{(1)}=E^{(3)}=E^{(6)}=E^{(8)}=J/2$, $E^{(2)}=E^{(5)}=0$
and $E^{(4)}=E^{(7)}=-J$ (see Appendix \ref{appa} with $B=0$). However, in this case the method cannot be implemented.
The degeneracy of the $E^{(2)}$ and $E^{(5)}$ egenvalues makes it impossible to measure the central spin. So, in the case of the absence
of a magnetic field the system achieves the entangled state during the minimal value of time but we cannot measure this state using the above method.
Let us consider our methods for preparation the maximally entangled states on the nuclear spins of xenon difluoride.

\section{Application to the xenon difluoride \label{sec5}}

\begin{figure}
\includegraphics[width=1.00\textwidth]{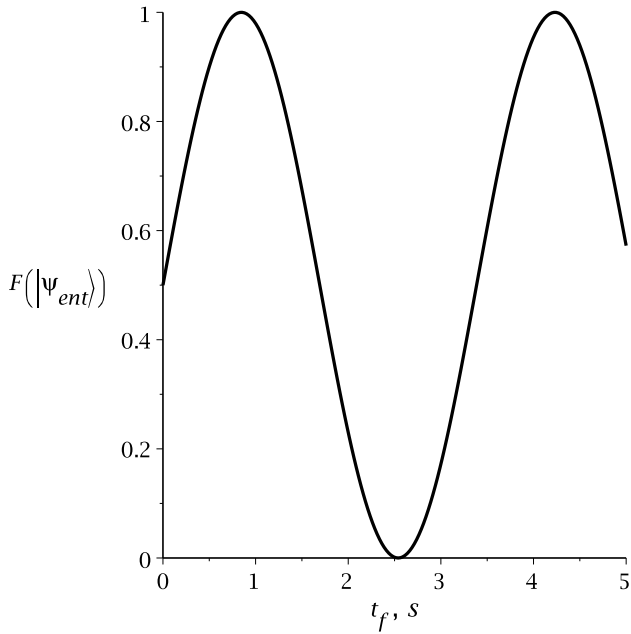}
\caption{Fidelity as a function of time (\ref{fidelityent}) between states (\ref{timedepstate2}) and (\ref{stateonps}) with $\vert z_{ent}\vert= 1$.
The results are obtained for nuclear spins of xenon difluoride placed in the magnetic field $B=1$ T.}
\label{fidelity}
\end{figure}

We propose to consider our method on nuclear spins of xenon difluoride. Xenon difluoride is a linear molecule (Fig.~\ref{model}) with a Xe atom which
is located between two atoms of F. We consider the system which consists of $^{129}$Xe and $^{19}$F isotopes because each atom of this molecule
has the nucleus with spin $1/2$. So, we assume that the nuclear spin of xenon atom plays the role of auxiliary spin and
the nuclear spins of fluorine atoms play the role of basic spins, respectively. The gyromagnetic ratios of nuclear spins of these isotopes are equal to
$\gamma_c=-73.997$ rad MHz T$^{-1}$ for $^{129}$Xe and $\gamma=251.662$ rad MHz T$^{-1}$ for $^{19}$F, respectively. The interaction
between nuclear spins is described by isotropic Heisenberg Hamiltonian. So, the orientation of molecule with respect to the magnetic field
does not impact the total Hamiltonian of system. The interaction between xenon and fluorine nuclear spins is much stronger than between
fluorine nuclear spins. Therefore, we can neglect the interaction between spins that
belong to the fluorine atoms. Thus, the evolution of system depends only on the interactions between fluorine and xenon nuclear spins. The interaction
couplings between these spins in different compounds, solvents and at certain temperatures were obtained in paper \cite{holloway1980}.
We use the interaction coupling between nuclear spins of $^{129}$Xe and $^{19}$F in xenon difluoride that is surrounded by BrF$_5$ molecules
as a solvent at temperature $-40$ $^{\circ}$C. This coupling is equal to $J/(2\pi\hbar)=5583$ Hz. Finally, it should be noted that the
long coherence time $>1$ s (see, for example \cite{Buluta2011}) of the nuclear spins is sufficient for implementation of quantum calculations on them.

As an example, let us consider the preparation of maximally entangled state on the spins of fluorine nuclei. So, for this purpose the system should be prepared
in the initial state $\vert\uparrow\uparrow\downarrow\rangle$, where the first ket vector corresponds to the state of the xenon nuclear spin and
the remaining vectors define the state of the fluorine nuclear spins. Then, we put the system in the external magnetic field $B$ directed along the $z$-axis.
At the moment of time $t_{ent}$ determined by condition (\ref{cocncurrencecond}) with $n=0$ we switch on the monochromatic electromagnetic radiation
with frequency $\omega=B\gamma_c$ and polarization vector with $z$-component which propagates along the $x$-axis. This field allows us to measure the
state of the system. So, if the integral intensity of the system is defined by equation (\ref{intensity}) with $\vert e_z\vert=1$ then
the state of xenon spin takes the form $\vert\uparrow\rangle$ and the fluorine spins with modulo a global phase achieve the maximally entangled state
\begin{eqnarray}
&&\vert\psi_{ent}\rangle=\frac{1}{\sqrt{2(1+\vert z_{ent}\vert^2)}}\nonumber\\
&&\times\left[\left(\vert z_{ent}\vert -i\right)\vert\uparrow\downarrow\rangle+\left(\vert z_{ent}\vert +i\right)\vert\downarrow\uparrow\rangle\right].
\label{stateonps}
\end{eqnarray}
Let us assume that the external magnetic field has the value $B=1$ T and is directed along the positive direction of the $z$-axis. The value of field is
sufficiently high to use equation (\ref{cocncurrencecond2}) for calculations and to put $\vert z_{ent}\vert\approx 1$.
Then we obtain $t_{ent}\simeq 0.82$ s. Finally, let us calculate the fidelity as a function of time between states (\ref{timedepstate2}) and (\ref{stateonps}). It takes the following form
\begin{eqnarray}
&&F(\vert\psi_{ent}\rangle)=\vert\langle\psi_{ent}\vert\psi_f\rangle\vert^2\nonumber\\
&&=\frac{1}{2}\left[1+\vert z_f\vert\sin\left(\phi_f+\frac{Jt_f}{4\hbar}+\frac{B(\gamma_c-\gamma)t_f}{2}\right)\right].
\label{fidelityent}
\end{eqnarray}
In Fig. \ref{fidelity}, we express the behavior of this expression. As we can see, the time which the system spends to achieve the maximally entangled
state is close to the coherence time of this system. This fact provides implementation of quantum calculations on such a system with high fidelity.

\section{Conclusions \label{sec6}}

We proposed the method for preparation of a two-qubit state on two basic spins-$1/2$ which mutually interact through an auxiliary spin.
We assumed that interaction between spins is described by the isotropic Heisenberg Hamiltonian and the basic spins are different
from the auxiliary spin by the gyromagnetic ratios. The essence of the method is that, initially, the three spins having started from the
factorized state evolve under the action of the external magnetic radiation during the predefined period of time. Then, using
the monochromatic electromagnetic field the auxiliary spin is measured. Depending on the result of measurement of predefined state (\ref{timedepstate2})
of the remaining spins, defined by the period of evolution and value of the magnetic field, is obtained with certain probability.
Using this probability the integral intensity was calculated (\ref{intensity}). This intensity allows to determine experimentally
the state of the system. Also, using the squared concurrence as a measure of entanglement, condition (\ref{cocncurrencecond})
for achieving the maximally entangled state was obtained. Finally, we described the process of the implementation of our method on the physical system
of nuclear spins of xenon difluoride. As an example, the conditions for preparing the maximally entangled state on this system were obtained.
It is important to note that the time which the system spends to achieve the required state is close to the coherence time of this system.
This fact allows to use the nuclear spins of xenon difluoride for implementation of quantum calculations with high fidelity.

\section{Acknowledgments}
The author thanks Profs. Volodymyr Tkachuk, Andrij Rovenchak, Drs. Volodymyr Pastukhov, Vasyl Vasyuta and Taras Verkholyak for useful comments.
This work was supported by Project FF-83F (No.~0119U002203) from the Ministry of Education and Science of Ukraine.

\begin{appendices}
\section{Eigenvalues and eigenstates of spin system \label{appa}}
\setcounter{equation}{0}
\renewcommand{\theequation}{A\arabic{equation}}

In this appendix, we present the eigenvalues and eigenstates of the three-spin system described by Hamiltonian (\ref{sysham2}). So, using the properties of the total spin-squared operator
${\bf S}^2$ and the ${\bf S}_{12}^2$ operator we obtain the following eigenvalues and eigenstates of this system
\begin{eqnarray}
&&E^{(1)}=\frac{J}{2}\left(\frac{B}{J}\hbar\left(\gamma_c+2\gamma\right)+1\right),\quad \vert\psi^{(1)}\rangle=\vert\uparrow\uparrow\uparrow\rangle, \nonumber\\
&&E^{(2)}=\frac{B\hbar\gamma_c}{2},\quad \vert\psi^{(2)}\rangle=\frac{1}{\sqrt{2}}\left(\vert\uparrow\uparrow\downarrow\rangle-\vert\uparrow\downarrow\uparrow\rangle\right),\nonumber\\
&&E^{(3)}=\frac{J}{2}\left(\frac{B}{J}\hbar\gamma+\frac{a_+-a_--1}{2}\right),\nonumber\\
&&\vert\psi^{(3)}\rangle=\frac{1}{\sqrt{a_+(a_+-a_-)}}\left(\vert\uparrow\uparrow\downarrow\rangle+\vert\uparrow\downarrow\uparrow\rangle+a_+\vert\downarrow\uparrow\uparrow\rangle\right),\nonumber\\
&&E^{(4)}=\frac{J}{2}\left(\frac{B}{J}\hbar\gamma-\frac{a_+-a_-+1}{2}\right),\nonumber\\
&&\vert\psi^{(4)}\rangle=\frac{1}{\sqrt{a_-(a_--a_+)}}\left(\vert\uparrow\uparrow\downarrow\rangle+\vert\uparrow\downarrow\uparrow\rangle+a_-\vert\downarrow\uparrow\uparrow\rangle\right),\nonumber\\
&&E^{(5)}=-\frac{B\hbar\gamma_c}{2},\quad \vert\psi^{(5)}\rangle=\frac{1}{\sqrt{2}}\left(\vert\downarrow\uparrow\downarrow\rangle-\vert\downarrow\downarrow\uparrow\rangle\right),\nonumber\\
&&E^{(6)}=-\frac{J}{2}\left(\frac{B}{J}\hbar\gamma-\frac{b_+-b_--1}{2}\right),\nonumber\\
&&\vert\psi^{(6)}\rangle=\frac{1}{\sqrt{b_+(b_+-b_-)}}\left(\vert\downarrow\uparrow\downarrow\rangle+\vert\downarrow\downarrow\uparrow\rangle+b_+\vert\uparrow\downarrow\downarrow\rangle\right),\nonumber\\
&&E^{(7)}=-\frac{J}{2}\left(\frac{B}{J}\hbar\gamma+\frac{b_+-b_-+1}{2}\right),\nonumber\\
&&\vert\psi^{(7)}\rangle=\frac{1}{\sqrt{b_-(b_--b_+)}}\left(\vert\downarrow\uparrow\downarrow\rangle+\vert\downarrow\downarrow\uparrow\rangle+b_-\vert\uparrow\downarrow\downarrow\rangle\right),\nonumber\\
&&E^{(8)}=-\frac{J}{2}\left(\frac{B}{J}\hbar\left(\gamma_c+2\gamma\right)-1\right),\quad \vert\psi^{(8)}\rangle=\vert\downarrow\downarrow\downarrow\rangle,\nonumber\\
\label{hamevandes}
\end{eqnarray}
where
\begin{eqnarray}
&&a_{\pm}=\pm\sqrt{\frac{9}{4}+\frac{B\hbar(\gamma_c-\gamma)}{J}\left(\frac{B\hbar(\gamma_c-\gamma)}{J}+1\right)}\nonumber\\
&&-\frac{B\hbar(\gamma_c-\gamma)}{J}-\frac{1}{2},\nonumber\\
&&b_{\pm}=\pm\sqrt{\frac{9}{4}+\frac{B\hbar(\gamma_c-\gamma)}{J}\left(\frac{B\hbar(\gamma_c-\gamma)}{J}-1\right)}\nonumber\\
&&+\frac{B\hbar(\gamma_c-\gamma)}{J}-\frac{1}{2}
\label{dimlessval}
\end{eqnarray}
are some dimensionless values.

\section{Derivation of the probabilities of obtaining the auxiliary spin in state $\vert\uparrow\rangle$ \label{appb}}
\setcounter{equation}{0}
\renewcommand{\theequation}{B\arabic{equation}}

During the process of measurement of the auxiliary spin, it can emit and absorb a photon. Therefore, two probabilities
which correspond to these events should be calculated. For this purpose, we use the time-dependent perturbation theory in the case of $H_{int}$ (\ref{haminteraction})
perturbation. At the moment of time $t_f$ we calculate in the first-order the corrections to the state $\vert\psi^{(5)}\rangle\vert N\rangle$,
where $\vert N\rangle$ is the eigenstate of photon subsystem. We calculate the corrections to this state because they define the auxiliary spin
in the state $\vert\uparrow\rangle$. So, in the first order of perturbation theory we obtain two states which correspond to the facts that
the photons are emitted and absorbed by the auxiliary spin. These states have the following form
\begin{eqnarray}
&&\vert\psi^{(5)+}\rangle=\vert\psi^{(5)}\rangle\vert N\rangle\nonumber\\
&&+\frac{2}{i\hbar}e^{i\omega_{52}\left(t_f+\tau/2\right)}V_{25}^+\frac{\sin\left(\omega_{52}\tau/2\right)}{\omega_{52}}\vert\psi^{(2)}\rangle\vert N+1\rangle,\nonumber\\
&&\vert\psi^{(5)-}\rangle=\vert\psi^{(5)}\rangle\vert N\rangle\nonumber\\
&&+\frac{2}{i\hbar}e^{i\omega_{25}\left(t_f+\tau/2\right)}V_{25}^-\frac{\sin\left(\omega_{25}\tau/2\right)}{\omega_{25}}\vert\psi^{(2)}\rangle\vert N-1\rangle,
\label{perturbedstate}
\end{eqnarray}
where
\begin{eqnarray}
&&V_{25}^+=\langle N+1\vert\langle\psi^{(2)}\vert H_{int}\vert\psi^{(5)}\rangle\vert N\rangle \nonumber\\
&&=\frac{1}{2}B_{int}\hbar\gamma_c e_ze^{-ikx}\sqrt{N+1},\nonumber\\
&&V_{25}^-=\langle N-1\vert\langle\psi^{(2)}\vert H_{int}\vert\psi^{(5)}\rangle\vert N\rangle \nonumber\\
&&=-\frac{1}{2}B_{int}\hbar\gamma_c e_ze^{ikx}\sqrt{N}.\nonumber
\label{pertcorr}
\end{eqnarray}
Finally, using the measurement postulate for state (\ref{timedepstate}) the probabilities, which correspond to the case that during the period of interaction
of the system with electromagnetic field the auxiliary spin takes the state $\vert\uparrow\rangle$, are calculated as follows
\begin{eqnarray}
&&W^{+}=\vert\langle N+1\vert \langle \psi(t)\vert\psi^{(5)+}\rangle\vert^2,\nonumber\\
&&W^{-}=\vert\langle N-1\vert \langle \psi(t)\vert\psi^{(5)-}\rangle\vert^2.
\label{pmprobab}
\end{eqnarray}
Calculating the corresponding scalar products and making some simplifications, we obtain expressions (\ref{probab}).

\end{appendices}


\begin{thebibliography}{99}
\bibitem{Kitaev2002} A. Yu. Kitaev, A. H. Shen, M. N. Vyalyi, {\it Classical and Quantum Computation} (American Mathematical Society, USA, 2002).
\bibitem{Krokhmalskii2004} T. Krokhmalskii, J. Phys. Stud. {\bf 8}, 1 (2004).
\bibitem{Barenco1995} A. Barenco, C. H. Bennett, R. Cleve, D. P. DiVincenzo, N. Margolus, P. Shor, T. Sleator, J. A. Smolin, H. Weinfurter,
Phys. Rev. A {\bf 52}, 3457 (1995).
\bibitem{qdots1} D. Loss, D. P. DiVincenzo, Phys. Rev. A \textbf{57}, 120 (1998).
\bibitem{qdots2} G. Burkard, D. Loss, D. P. DiVincenzo, Phys. Rev. B \textbf{59}, 2070 (1999).
\bibitem{qdots3} N. Rohling, G. Burkard, Phys. Rev. B \textbf{88}, 085402 (2013).
\bibitem{phosphorus3} B. E. Kane, Nature \textbf{393}, 133 (1998).
\bibitem{scc} J. R. Petta, A. C. Johnson, J. M. Taylor, E. A. Laird, A. Yacoby, M. D. Lukin, C. M. Marcus, M. P. Hanson and A.C. Gossard,
Science \textbf{309}, 2180 (2005).
\bibitem{semcond1} S. Kuhlen, K. Schmalbuch, M. Hagedorn, P. Schlammes, M. Patt, M. Lepsa, G. G\"untherodt and B. Beschoten, Phys. Rev. Lett. \textbf{109},
146603 (2012).
\bibitem{semcond2} R. Vrijen, E. Yablonovitch, K. Wang, H. W. Jiang, A. Balandin, V. Roychowdhury, T. Mor, D. DiVincenzo,
Phys. Rev. A \textbf{62}, 012306 (2000).
\bibitem{semcond3} H. Bluhm, S. Foletti, I. Neder, M. Rudner, D. Mahalu, V. Umansky, A. Yacoby, Nature Physics \textbf{7}, 109 (2011).
\bibitem{semcond4} F. H. L. Koppens, C. Buizert, K. J. Tielrooij, I. T. Vink, K. C. Nowack, T. Meunier, L. P. Kouwenhoven and L. M. K. Vandersypen,
Nature \textbf{442}, 766 (2006).
\bibitem{semcond5} K. C. Nowack, F. H. L. Koppens, Yu. V. Nazarov and L. M. K. Vandersypen, Science \textbf{318}, 1430 (2007).
\bibitem{semcond6} B. M. Maune et al., Nature \textbf{481}, 344 (2012).
\bibitem{phosphorus1} J. J. Pla, K. Y. Tan, J. P. Dehollain, W. H. Lim, J. J. L. Morton, F. A. Zwanenburg, D. N. Jamieson,
A. S. Dzurak, A. Morello, Nature \textbf{496}, 334 (2013).
\bibitem{phosphorus2} J. J. Pla, K. Y. Tan, J. P. Dehollain, W. H. Lim, J. J. L. Morton, D. N. Jamieson, A. S. Dzurak, Andrea Morello,
Nature \textbf{489}, 541 (2012).
\bibitem{phosphorus4} A. Morello et al., Nature \textbf{467}, 687 (2010).
\bibitem{phosphorus6} A. M. Tyryshkin et al., Nature Materials \textbf{11}, 143 (2012).
\bibitem{phosphorus7} J. J. L. Morton, A. M. Tyryshkin, R. M. Brown, Sh. Shankar, B. W. Lovett, A. Ardavan, T. Schenkel, E. E. Haller, J. W. Ager,
S. A. Lyon, Nature \textbf{455}, 1085 (2008).
\bibitem{phosphorus8} M. Steger, K. Saeedi, M. L. W. Thewalt, J. J. L. Morton, H. Riemann, N. V. Abrosimov, P. Becker and H. J. Pohl,
Science \textbf{336}, 1280 (2012).
\bibitem{HinSi} A. J. Skinner, M. E. Davenport and B. E. Kane, Phys. Rev. Lett. \textbf{90}, 087901 (2003).
\bibitem{BiinSi} M. H. Mohammady, G. W. Morley, A. Nazir and T. S. Monteiro, Phys. Rev. B \textbf{85}, 094404 (2012).
\bibitem{LNT} K. Saeedi, S. Simmons, J. Z. Salvail, P. Dluhy, H. Riemann, N. V. Abrosimov, P. Becker, H.-J. Pohl,
J. J. L. Morton, M. L. W. Thewalt, Science \textbf{342}, 830 (2013).
\bibitem{Christle2017} D. J. Christle et al, Phys. Rev. X {\bf 7}, 021046 (2017).
\bibitem{Nichol2017} J. M. Nichol, L. A. Orona, Sh. P. Harvey, S. Fallahi, G. C. Gardner, M. J. Manfra, A. Yacoby, npj Quantum Information {\bf 3}, 3 (2017).
\bibitem{Harvey-Collard2018} P. Harvey-Collard et al, Phys. Rev. X {\bf 8}, 021046 (2018).
\bibitem{Nagy2019} R. Nagy et al, Nature Communications {\bf 10}, 1954 (2019).
\bibitem{supcond1} L. F. Wei, Yu-xi Liu and Franco Nori, Phys. Rev. B \textbf{71}, 134506 (2005).
\bibitem{supcond2} J. E. Mooij, T. P. Orlando, L. Levitov, Lin Tian, Caspar H. van der Wal  and Seth Lloyd, Science \textbf{285}, 1036 (1999).
\bibitem{supcond3} Y. Makhlin, G. Sch\"on, A. Shnirman, Rev. Mod. Phys. \textbf{73}, 357 (2001).
\bibitem{supcond4} J. Majer et al., Nature \textbf{449}, 443 (2007).
\bibitem{SchrodCat1} K. Molmer, A. Sorensen, Phys. Rev. Lett. {\bf 82}, 1835 (1999).
\bibitem{EQSSTI} D. Porras, J. I. Cirac, Phys. Rev. Lett. {\bf 92}, 207901 (2004).
\bibitem{ITQLLWR} F. Mintert, Ch. Wunderlich, Phys. Rev. Lett. {\bf 87}, 257904 (2001).
\bibitem{SchrodCat2} D. Leibried etc, Nature {\bf 438}, 639 (2005).
\bibitem{ETDIITIQSHI} J. W. Britton, B. C. Sawyer, A. C. Keith, C. C. Joseph Wang, J. K. Freericks, H. Uys, M. J. Biercuk, J. J. Bollinger,
Nature {\bf 484}, 489 (2012).
\bibitem{QSDEGHTI} J. G. Bohnet, B. C. Sawyer, J. W. Britton, M. L. Wall, A. M. Rey, M. Foss-Feig, J. J. Bollinger,
Science {\bf 352}, 1297 (2016).
\bibitem{QSWTI} R. Blatt, C. F. Roos, Nature Physics {\bf 8}, 277 (2012).
\bibitem{opticallattice1} L.-M. Duan, E. Demler, M. D. Lukin, Phys. Rev. Lett. {\bf 91}, 090402 (2003).
\bibitem{opticallattice18} A. B. Kuklov, B. V. Svistunov, Phys. Rev. Lett. {\bf 90}, 100401 (2003).
\bibitem{opticallattice5} I. Bloch, {\it Many-Body Physics with Ultracold Gases Edited by C. Salomon, G. Shlyapnikov, L. F. Cugliandolo}
(Oxford University Press, Oxford, UK, 2013), pp. 71-108.
\bibitem{opticallattice19} Ch. Gross, I. Bloch, Science {\bf 357}, 995 (2017).
\bibitem{opticallattice6} D. Jaksch, Contemporary Physics {\bf 45}, 367 (2004).
\bibitem{opticallattice7} I. Bloch, Nature {\bf 453}, 1016 (2008).
\bibitem{opticallattice9} J. I. Cirac, P. Zoller, Phys. Rev. Lett. {\bf 74}, 4091 (1995).
\bibitem{Ladd2010} T. D. Ladd, F. Jelezko, R. Laflamme, Y. Nakamura, C. Monroe, J. L. O'Brien, Nature {\bf 464}, 45 (2010).
\bibitem{Buluta2011} I. Buluta, S. Ashhab, F. Nori, Rep. Prog. Phys. {\bf 74}, 104401 (2011).
\bibitem{srtech} C. P. Slichter, {\it Principles of Magnetic Resonance} (Springer-Verlag, Berlin, 1990).
\bibitem{Vandersypen2004} L. M. K. Vandersypen, I. L. Chuang, Rev. Mod. Phys. {\bf 76}, 1037 (2004).
\bibitem{Mehring2003} M. Mehring, J. Mende, W. Scherer, Phys. Rev. Lett. {\bf 90}, 153001 (2003).
\bibitem{Kuzmak2014} A .R. Kuzmak, V. M. Tkachuk, Phys. Lett. A {\bf 378}, 1469 (2014).
\bibitem{Kuzmak2018} A .R. Kuzmak, Int. J. Quan. Inf. {\bf 16}, 1850044 (2018).
\bibitem{Sahling2015} S. Sahling et al., Nature Physics {\bf 11}, 255 (2015).
\bibitem{twosqg1} A. Carlini, A. Hosoya, T. Koike and Y. Okudaira, Phys. Rev. A \textbf{75}, 042308 (2007).
\bibitem{twosqg2} A. R. Kuzmak, V. M. Tkachuk, J. Phys. A \textbf{46}, 155305 (2013).
\bibitem{twosqg3} N. Khaneja, S. J. Glaser and R. Brockett, Phys. Rev. A \textbf{65}, 032301 (2002).
\bibitem{twosqg4} T. O. Reiss, N. Khaneja, S. J. and Glaser, J. Magn. Reson \textbf{165}, 95 (2003).
\bibitem{twosqg5} H. Yuan, N. Khaneja, Phys. Rev. A \textbf{72}, 040301(R) (2005).
\bibitem{twosqg6} R. Zeier, H. Yuan and N. Khaneja, Phys. Rev. A \textbf{77}, 032332 (2008).
\bibitem{twosqg7} N. Schuch and J. Siewert, Phys. Rev. A \textbf{67}, 032301 (2003).
\bibitem{Zu2014} C. Zu, W.-B. Wang, L. He, w.-G. Zhang, C.-Y. Dai, F. Wang, L.-M. Duan, Nature {\bf 514}, 72 (2014).
\bibitem{Ashhab2012} S. Ashhab, P. C. de Groot, Franco Nori, Phys. Rev. A {\bf 85}, 052327 (2012).
\bibitem{bennet1993} C. H. Bennett, G. Brassard, C. Crepeau, R. Jozsa, A. Peres, W. K. Wootters, Phys. Rev. Lett. {\bf 70}, 1895 (1993).
\bibitem{zeilinger1997} D. Bouwmeester, Jian-Wei Pan, K. Mattle, M. Eibl, H. Weinfurter and A. Zeilinger, Nature {\bf 390}, 575 (1997).
\bibitem{giovannetti20031} V. Giovannetti, S. Lloyd, L. Maccone, Europhys. Lett. \textbf{62}, 615 (2003).
\bibitem{giovannetti20032} V. Giovannetti, S. Lloyd, L. Maccone, Phys. Rev. A \textbf{67}, 052109 (2003).
\bibitem{zander2007} C. Zander, A. R. Plastino, A. Plastino, M. Casas, J. Phys. A \textbf{40}, 2861 (2007).
\bibitem{borras20081} A. Borras, C. Zander, A. R. Plastino, M. Casas, A. Plastino, Europhys. Lett. \textbf{81}, 30007 (2008).
\bibitem{borras20082} A. Borras, A. R. Plastino, M. Casas, A. Plastino, Phys. Rev. A \textbf{78}, 052104 (2008).
\bibitem{zhao2009} Bao-Kui Zhao, Fu-Guo Deng, Feng-Shou Zhang, Hong-Yu Zhou, Phys. Rev. A \textbf{80}, 052106 (2009).
\bibitem{Woo} W. K. Wootters, Quan. Inf. Comp. {\bf 1}, 27 (2001).
\bibitem{ent1} W. K. Wootters, Phys. Rev. Lett. {\bf 80}, 2245 (1998).
\bibitem{ent2} S. Hill, W. K. Wootters, Phys. Rev. Lett. {\bf 78}, 5022 (1997).
\bibitem{holloway1980} J. H. Holloway, G. J. Schrobilgen, Inorg. Chem. {\bf 19}, 2632 (1980).
\end{thebibliography}
\end{document}